\documentclass{PoS}

\title{Scalar-pseudoscalar meson spectrum in SU(3) PNJL model}

\ShortTitle{Scalar-pseudoscalar meson spectrum in SU(3) PNJL model }

\author{\speaker{P. Costa}\\%
        Departamento de Física, Universidade de Coimbra, Portugal\\
        E.S.T.G., Instituto Politécnico de Leiria, Morro do Lena-Alto do Vieiro, 2411-901 Leiria, Portugal\\
E-mail: \email{pcosta@teor.fis.uc.pt}}

\author{ M. C. Ruivo\\
Departamento de Física, Universidade de Coimbra, Portugal\\
E-mail: \email{maria@teor.fis.uc.pt}}

\author{ C. A. de Sousa\\
Departamento de Física, Universidade de Coimbra, Portugal\\
E-mail: \email{celia@teor.fis.uc.pt}}

\author{ H. Hansen\\
Univ.Lyon/UCBL, CNRS/IN2P3, IPNL, 69622
Villeurbanne Cedex, France\\
E-mail: \email{hansen@ipnl.in2p3.fr}}

\author{ W.M. Alberico\\
Dipartimento di Fisica Teorica, University of Torino and INFN, Sezione di Torino, via P.
Giuria~1, I-10125 Torino, Italy \\
E-mail: \email{alberico@to.infn.it}}

\abstract{We investigate the modifications  of mesonic properties  in a hot medium
having in mind to use the mesonic modes as a tool  to analyze the restoration of chiral and axial
symmetries in the context of the SU(3) Polyakov--Nambu--Jona-Lasinio (PNJL) model. The
results lead to the conclusion that the effects of the Polyakov loop are fundamental for
reproducing lattice findings. It is found that the restoration of chiral symmetry in the PNJL model
occurs in a small range of temperature (smaller that in the Nambu--Jona-Lasinio (NJL)), and 
the topological susceptibility reproduces well the lattice results around $T/T_c^\chi\approx 1.0$.}

\FullConference{8th Conference Quark Confinement and the Hadron Spectrum\\
         September 1-6, 2008\\
         Mainz. Germany}

\begin{document}


The study of the modifications of particles propagating in a hot or dense matter is an
interesting tool to investigate the restoration of chiral and axial symmetries.
Understanding the restoration of symmetries and deconfinement, which are expected to
occur under extreme conditions (high density and/or temperature), and, more recently, the
possible survival of bound states in the deconfined phase of QCD, is a challenging
problem \cite{datta1,albe}.

In Ref. \cite{Hansen:2006PRD} the study of meson properties, in the SU(2) sector, around
the critical region, was performed in the framework of the modified Nambu--Jona-Lasinio
model including the Polyakov loop (the so-called PNJL model)
\cite{Meisinger:1995ih,Megias:2006PRD,Ratti:2005jh}.
The inclusion of the Polyakov loop is fundamental for reproducing lattice data concerning
QCD thermodynamics \cite{Ratti:2005jh}, once it originates a suppression of the unwanted
quark contributions to the thermodynamics below the critical temperature.

In this paper, we  generalize the investigation  to the SU(3) sector. Since the scalar
mesons ($\sigma,\,f_0,\,a_0$ and $K^*_0$) can be considered as chiral partners of the
pseudoscalar nonet ($\eta,\eta^\prime,\pi^0$ and $K$), it is interesting to see how the
Polyakov loop affects the behavior of these mesons, having in mind the investigation of
effective restoration of chiral symmetry.

The flavor mixing induced by the presence of the axial anomaly causes a violation of the
OZI rule, both for scalar and pseudoscalar mesons, hence the restoration  of axial
symmetry should have relevant consequences on the phenomenology of the mesonic mixing
angles,  as well as on the topological susceptibility.

We perform our calculations in the framework of an extended SU(3) PNJL Lagrangian, which
contains the 't Hooft instanton induced interaction term that breaks the U$_A$(1)
symmetry  \cite{Fukushima:2008PRD}:
\begin{eqnarray}
{\mathcal L_{PNJL}\,}&=& \bar q(i \gamma^\mu D_\mu-\hat m)q\nonumber
\,+\,\frac{1}{2}\,g_S\,\,\sum_{a=0}^8\, [\,{(\,\bar q\,\lambda^a\, q\,)}
^2\,\,+\,\,{(\,\bar q \,i\,\gamma_5\,\lambda^a\, q\,)}^2\,] \nonumber\\
&+& g_D\,\{\mbox{det}\,[\bar q\,(1+\gamma_5)\,q] +\mbox{det} \,[\bar
q\,(1-\gamma_5)\,q]\} \,-\, \mathcal{U}\left(\Phi,\bar\Phi;T\right). \label{eq:lag}
\end{eqnarray}

The Polyakov loop is an order parameter for the restoration of the ${\mathbb Z}_3$ (the
center of SU$_c$(3)) symmetry of QCD and is related to the deconfinement phase
transition: ${\mathbb Z}_3$ is broken in the deconfined phase ($\Phi \rightarrow 1$) and
restored in the confined one ($\Phi \rightarrow 0$) \cite{Polyakov}.

\begin{figure}[t]
  \begin{center}
    \hspace{-0.5cm}
    \includegraphics[width=0.65\textwidth]{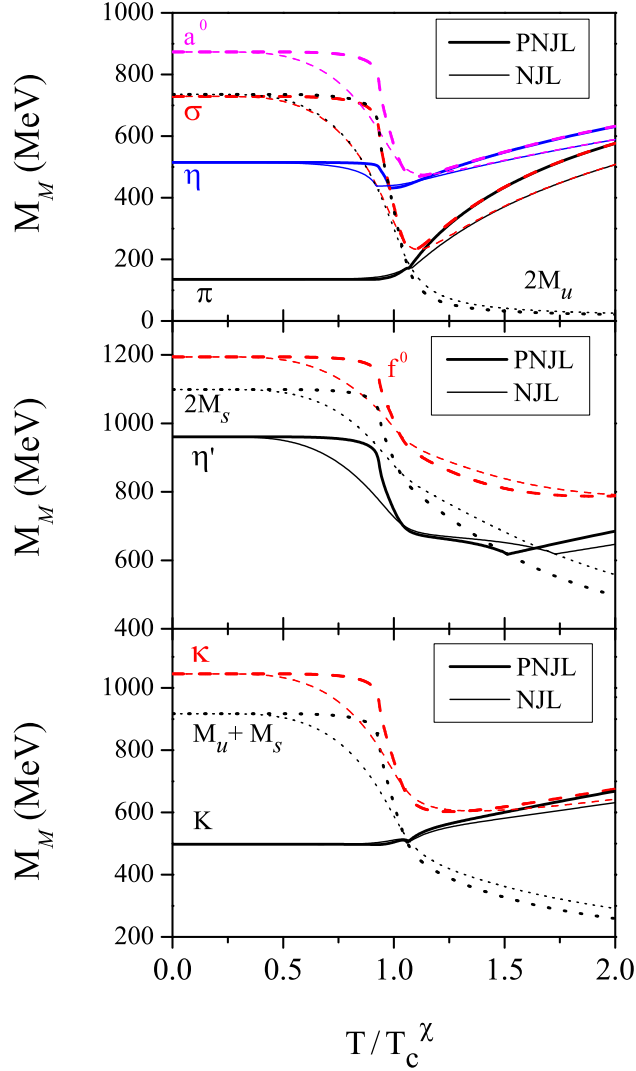} 
      \caption{\label{fig:Mmesons} 
	      Comparison of the pseudoscalar and scalar meson
        masses in PNJL (thick lines) and NJL (thin lines) models as functions
        of the reduced temperature $T/T_c^{\chi}$. In the upper panel
        the $a_0$ (dashed line), $\sigma$ (dot-dashed line), $\eta$
        and $\pi$ (continuous lines) are shown, together with 
        $2M_u$ (dotted lines). In the middle panel $f_0$ (dashed
        line) and $\eta^{\prime}$ (continuous line) are shown and
        compared with $2M_s$ (dotted line). In the lower
        panel the $\kappa$ (dashed line) and $K$ (continuous line)
        masses are compared with $M_u+M_s$ (dotted line).}
  \end{center}
\end{figure}

We start our study by analyzing the masses of the strange and non-strange quarks as
functions of the temperature (on Fig. \ref{fig:Mmesons} the dressed quark masses 
$2M_u$, $2M_s$ and $M_u+M_S$ are displayed). We verify that the mass of the light 
quarks drops to the current quark mass, indicating a smooth crossover from the chiral 
broken to an approximate chiral symmetric phase. This dropping is more pronounced in the 
PNJL model than in the NJL one. The strange quark mass presents a very similar behavior, with a
significant decrease above the chiral transition temperature $T_c^\chi$, however its mass
is still far away from the strange current quark mass. As usual chiral symmetry shows a
slow tendency to get restored in the strange sector, however this tendency is faster in
the PNJL model. 
As in the NJL model \cite{costa:PRD}, once $m_u=m_d<m_s$, the (sub)group SU(2)
$\otimes$SU(2) is a much better symmetry of the Lagrangian (\ref{eq:lag}).

Nevertheless, the fact that the masses of the quarks drop faster around $T_c^\chi$ in
PNJL model is important for the mesonic properties (for example it could modify the
survival of mesonic bound states in the plasma phase). Besides, due to the strangeness 
content of some mesons, the behavior of the strange quark (modified by the Polyakov loop)
is important for their properties, as well as for other observables related 
to the axial anomaly (as noticed in \cite{costa:2007PRD}, topological susceptibility 
is strongly influenced by the strange sector).

Let us now analyze the behavior of the mesons in connection with  possible restoration of
symmetries. It can be seen in Fig. \ref{fig:Mmesons} (upper panel) that the
partners ($\pi,\sigma$) and ($\eta,a_{0}$) become degenerate  at almost the same
temperature. In both models, this behavior indicates the effective restoration of
chiral symmetry in the non-strange sector.

On the contrary, the $\eta^{\prime}$ and $f_{0}$ masses (Fig. \ref{fig:Mmesons} middle panel) do not show 
a tendency to converge in the region of temperatures studied, a behavior that reflects 
the fact that chiral symmetry does not get restored in the strange sector. 
Finally, we focus on   the $ \kappa$-meson (lower panel): it is always an unbound, resonant state
and, as the temperature increases, it tends to become degenerate in mass with the
$K$-meson, but at larger temperatures (of the order of 1.5~$T_c^\chi$ in PNJL, and higher
in NJL). In summary, the masses of the $\sigma$ and $\eta$ (that become less strange as the
temperature increases and the mixing in the strange sector decreases)
converge, respectively, with those of the non-strange, $\pi^0$ and $a_0$. The
convergence of the chiral partners $\kappa$ and $K$ (Fig. \ref{fig:Mmesons} lower panel), 
that have a $\bar u s$ structure, occurs  at  higher temperatures and is probably slowed down by 
the  small decrease of the strange quark mass, $M_s$.

Concerning the axial symmetry, its effective restoration should be signaled by the
vanishing of the observables related to the anomaly, like  the gap between the masses of
the  chiral partners of the U$_A$(1) symmetry, and the topological susceptibility. For 
what concerns the observables here analyzed, we notice that although both models exhibit
a tendency to a decrease of the anomaly effects, the restoration of the axial symmetry is
not achieved: the masses of the partners ($\pi, \eta$) and ($\sigma, a_0$) although
getting close at high temperatures, do not converge. We also verify that   the mixing
angles do not reach the ideal values~\cite{costaW}. This is expected since, in the
framework of the NJL model, it has been shown that only with additional assumptions  (for
example,  by choosing a temperature dependent anomaly coefficient~\cite{costa:PRD} or
using a regularization where the cutoff goes to infinity at $T\not=0$
\cite{costa:2007PRD}) the restoration of the axial symmetry can be achieved.

We point out that the new feature
of the PNJL model is that the faster decrease of the quark  condensates  leads to a
faster partial restoration of chiral symmetry.  In addition, although the axial chiral
partners do not converge, in PNJL model the masses of ($\pi, \eta$) become closer than in
NJL, as well as those of ($\sigma, a_0$). This is an indication that, although axial
symmetry is not restored in the range of temperatures studied, the tendency to the
possible restoration of this symmetry is faster in the PNJL model.

\begin{figure}[t]
  \begin{center}
    \includegraphics[width=0.8\textwidth]{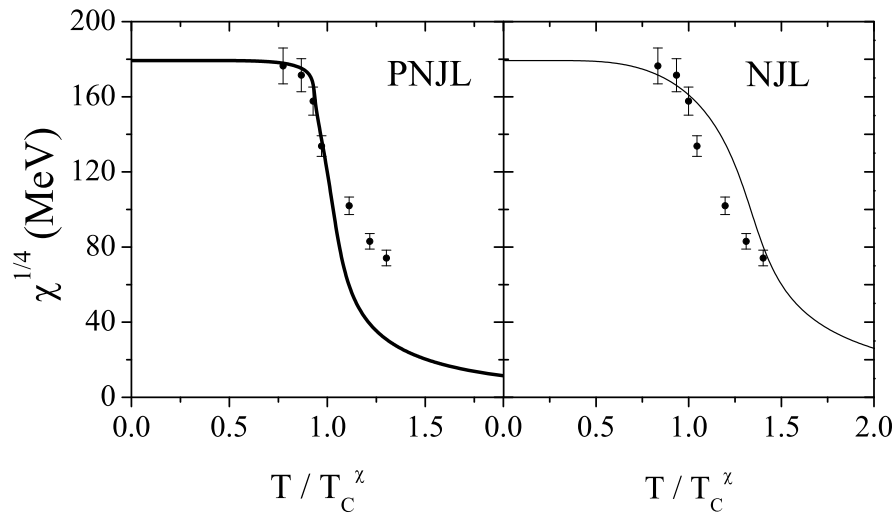} 
    \caption{\label{fig:TopSuscep}
    Topological susceptibility as a function of the reduced
    temperature $T/T_c^{\chi}$ for the PNJL (left panel) and the NJL (right panel) models.}
  \end{center}
\end{figure}

We also  derive the topological susceptibility, $\chi$ (Fig.~\ref{fig:TopSuscep}), 
that is  far away from being zero in both models. However,
it is interesting to notice that the PNJL calculation  nicely reproduces the first
lattice points, namely the rather steep drop around $T_c^\chi$, while this is not
verified in the NJL model. Although restoration of axial symmetry is not achieved, this
behavior of the topological susceptibility indicates that the PNJL model shows a more
pronounced tendency for the restoration  of the axial symmetry.

\begin{acknowledgments}
Work supported by Grant SFRH/BPD/23252/2005 (P. Costa) and by F.C.T. under projects
POCI/FP/63945/2005 and POCI/FP/81936/2007 (H. Hansen).
\end{acknowledgments}


\begin{thebibliography}{99}

\bibitem{datta1}
    S. Datta, F. Karsch, P. Petreczky and I. Wetzorke,
    Phys. Rev. D {\bf 69}, 094507 (2004).

\bibitem{albe}
    W.M. Alberico, A. Beraudo, A. De Pace and  A. Molinari,
    Phys. Rev. D {\bf 72}, 114011 (2005).

\bibitem{Hansen:2006PRD}
    H. Hansen, W. M. Alberico, A. Beraudo, A. Molinari, M. Nardi, and C. Ratti,
    Phys. Rev. D {\bf 75}, 065004 (2007).

\bibitem{Meisinger:1995ih}
    P.~N. Meisinger and M.~C. Ogilvie,
    Phys. Lett. B {\bf 379}, 163 (1996).

\bibitem{Megias:2006PRD}
    E. Megias, E. R. Arriola, and L.L. Salcedo,
    Phys. Rev. D {\bf 74}, (2006) 065005.

\bibitem{Ratti:2005jh}
    C. Ratti, M.~A. Thaler, and W. Weise,
    Phys. Rev. D {\bf 73},  014019  (2006).

\bibitem{Fukushima:2008PRD}
    Kenji Fukushima,
    Phys. Rev. D {\bf 77}, 114028 (2008).

\bibitem{Polyakov}
    A. M. Polyakov,
    Phys. Lett. B {\bf 72}, 477 (1978).

\bibitem{costa:PRD}
    P. Costa, M. C. Ruivo, C. A. de Sousa, and Y. L. Kalinovsky,
    Phys. Rev. D {\bf 70}, 116013 (2004);
    Phys. Rev. D {\bf 71}, 116002 (2005).

\bibitem{costaW}
    P. Costa, M. C. Ruivo, C. A de Sousa , H. Hansen, and W. M. Alberico, 
    arXiv:0807.2134 [hep-ph].

\bibitem{costa:2003PRC}
    P. Costa, M. C. Ruivo, C. A de Sousa and Yu. L. Kalinovsky,
    Phys. Rev. C {\bf 70}, 025204 (2004).

\bibitem{costa:2007PRD}
    P. Costa, M. C. Ruivo, and C. A de Sousa,
    Phys. Rev. D {\bf 77}, 096009 (2008).


\end{thebibliography}
\end{document}